\documentclass[a4paper]{jpconf}

\pdfoutput=1

\usepackage{iopams}
\usepackage{graphicx}
\usepackage{type1cm}

\begin{document}

\title{The effective kinetic term in CDT\footnote{Based on a talk given at {\it Loops `11} in Madrid, Spain on May 24th, 2011.}}

\author{T G Budd}
\address{Institute for Theoretical Physics, Utrecht University,  3508 TD, Utrecht, The Netherlands}

\ead{t.g.budd@uu.nl}

\begin{abstract}
We report on recently performed simulations of Causal Dynamical Triangulations (CDT) in 2+1 dimensions aimed at studying its effective dynamics in the continuum limit. Two pieces of evidence from completely different measurements are presented suggesting that three-dimensional CDT is effectively described by an action with kinetic term given by a modified Wheeler-De Witt metric. These observations could strengthen an earlier observed connection between CDT and Ho\v{r}ava-Lifshitz gravity. One piece of evidence comes from measurements of the modular parameter in CDT simulations with spatial topology of a torus, the other from measurements of local metric fluctuations close to a fixed spatial boundary. 
\end{abstract}

\section{Introduction}

Causal dynamical triangulation (CDT) provides an explicit realization of the sum over space-time geometries which appears in the path-integral approach to quantum gravity. Its simplicity allows us to feed the model into a computer and find expectation values of observables numerically. We attempt to exploit the simplicity to address one of the hardest challenges faced by non-perturbative approaches to quantum gravity: showing that at low energies a classical geometrical theory emerges. 

CDT has provided several pieces of evidence suggesting that it might possess a continuum limit showing classical behaviour. However, at present it is unclear what effective theory governs this classical regime. In this paper we describe two distinct sets of observables which probe the semi-classical regime of CDT in 2+1 dimensions and allow us to obtain non-trivial information about the character of the effective theory. Before describing these, let us first briefly introduce the set-up of CDT in 2+1 dimensions.

The motivation for CDT comes from the assumption that path integrals over Lorentzian geometries can be computed with the help of statistical path integrals over Euclidean geometries that are equipped with a time-foliation. The leaves of a foliated geometry are interpreted as spatial geometries and their topology is assumed not to change in time. In CDT the path integral is replaced by a discrete sum over all such geometries that can be assembled from a finite number of equilateral tetrahedra. More precisely we have the partition function
\begin{equation}\label{eq:partitionfunction}
Z_{\mathrm{CDT}} = \sum_{\mathrm{triangulations}\,T} e^{-S_{\mathrm{CDT}}[T]} \quad \mathrm{with} \quad S_{\mathrm{CDT}}[T] = k_3 N_3 - k_0 N_0.
\end{equation}
The action $S_{\mathrm{CDT}}$ is derived from the Regge action and depends linearly on the number $N_3$ of tetrahedra and the number $N_0$ of vertices in the triangulation $T$. The three-dimensional triangulations are required to be foliated in the sense that they can be constructed from a sequence of two-dimensional triangulations connected by slices of tetrahedra (see figure \ref{fig:extrinsic}).

We use Monte Carlo methods to randomly generate CDT configurations according to the Boltzmann distribution in (\ref{eq:partitionfunction}). By performing measurements on individual triangulations we can approximate expectation values of observables. A particularly simple observable is the volume $V(t)$ of the spatial triangulation at time $t$. It has been previously observed\footnote{See \cite{AJL} for results in 2+1 dimensions and \cite{AGJL} for more recent results in 3+1 dimensions.} that its expectation value $\langle V(t) \rangle$ for CDT with spatial topology of a 2-sphere corresponds to that of a proper-time slicing of a round 3-sphere, which is the Euclidean counterpart of de Sitter space. Both the expectation value and its quantum fluctuations turn out to be well described by an effective action
\begin{equation}\label{eq:effspherical}
S_{\mathrm{eff}}[V] = \int dt\left( c_0 \frac{\dot{V}^2}{V} - c_1 V\right) \quad \mathrm{with} \quad c_0,c_1 > 0,
\end{equation}
which is very similar to the (Euclidean) Einstein-Hilbert action $\int d^3x\sqrt{g}(-R +2\Lambda)$ evaluated on a spherical cosmology $ds^2 = dt^2 + V(t) d\Omega^2$. The only difference is an overall minus sign, which ensures that for fixed 3-volume (\ref{eq:effspherical}) is bounded below while the Einstein-Hilbert one is not.

This is a general property of the Euclidean Einstein-Hilbert action and is known as the conformal mode problem. We can illustrate this by writing the metric in proper-time gauge $ds^2 = dt^2 + g_{ab} dx^a dx^b$, in which case the Einstein-Hilbert action takes the form 
\begin{equation}\label{eq:admaction}
S_{\mathrm{EH}} = \kappa \int dt \int d^2x\sqrt{g} \left(\frac{1}{4} \dot{g}_{ab} \mathcal{G}^{abcd}\dot{g}_{cd} - R + 2\Lambda\right), \quad \mathcal{G}^{abcd} = \frac{1}{2}\left(g^{ac}g^{bd}+g^{ad}g^{bc}\right) - g^{ab}g^{cd},
\end{equation}
and $\mathcal{G}^{abcd}$ is referred to as the Wheeler-De Witt supermetric. This metric on the linear space of metric deformations is indefinite: it is positive definite on traceless deformations but negative definite on conformal deformations. As a consequence (\ref{eq:admaction}) is not a suitable ansatz for an effective action for CDT: we want such an effective action to describe, at least in a semi-classical way, the quantum fluctuations we observe in our simulations and to do that it needs to have a (local) minimum at the classical solutions.

Inspired by the Ho\v{r}ava-Lifshitz approach to gravity \cite{H} and the presence of a preferred time-slicing in CDT, we propose a modification of the ansatz (\ref{eq:admaction}). Namely we modify the Wheeler-De Witt metric by inserting a free parameter $\lambda$,
\begin{equation}\label{eq:modifiedWdW}
\mathcal{G}^{abcd} \;\;\to \;\;
\mathcal{G}^{abcd}_{\lambda} = \frac{1}{2}\left(g^{ac}g^{bd}+g^{ad}g^{bc}\right) - \lambda g^{ab}g^{cd}.
\end{equation}
This way we obtain the most general ultralocal kinetic term compatible with foliation-preserving diffeomorphism invariance. This metric is positive definite in the regime $\lambda < 1/2$, while in general relativity we have the value $\lambda = 1$.

In the remainder of this paper we will put this ansatz to test by measuring observables that are only sensitive to the kinetic term in the effective action.

\section{CDT on a torus}

\begin{figure}
\begin{center}
\begin{minipage}{7.5cm}
\includegraphics[width=7.5cm]{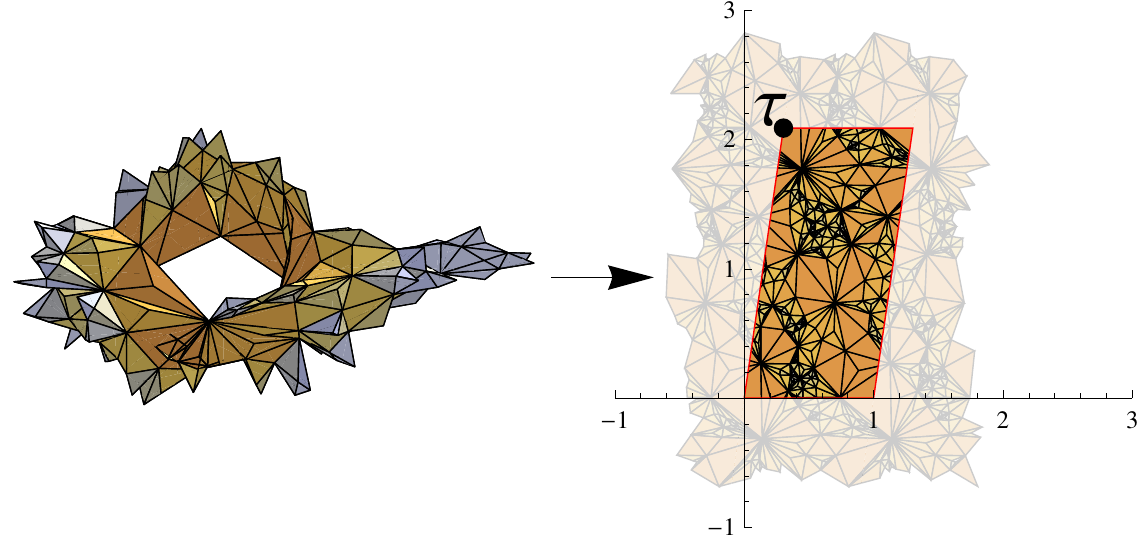}
\caption{A harmonic embedding of a triangulation of the torus. The modular parameter $\tau=\tau_1+i \tau_2$ is associated with the shape of the parallelogram in the manner shown.\label{fig:triangulation}}\end{minipage}
\hspace{0.5cm}
\begin{minipage}{7.5cm}
\includegraphics[width=7.2cm]{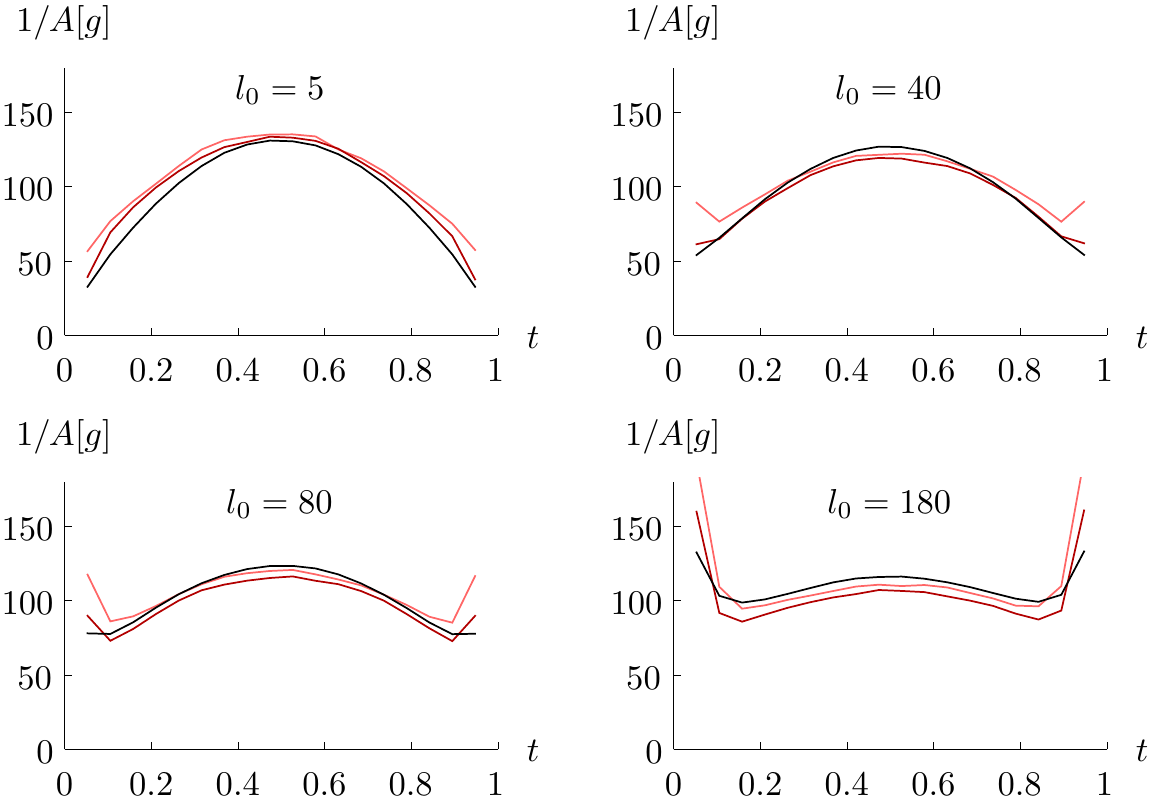}
\caption{The quantity $1/A[g]$ deduced from direct measurement (in black) and from the correlations of $\tau_1$ (in orange) and $\tau_2$ (in red). \label{fig:cinv}}
\end{minipage}
\end{center}
\end{figure}

We would like to see what consequence our ansatz has for cosmological models. Clearly the spatial volume is an observable which only depends on the conformal degrees of freedom in the metric. In order to probe the kinetic term in a non-trivial way, we need another observable that depends also on traceless degrees of freedom, i.e. an observable that measures the shape of space. Taking the spatial topology to be that of the torus, we have a prime candidate for such an observable: the complex modular parameter $\tau$ describing the conformal structure. It is well-known that any metric on the torus is conformally related to a unique flat metric of unit volume. Up to diffeomorphisms the latter form a 2-parameter family and can be obtained by gluing sides of a parallelogram in the Euclidean plane (see figure \ref{fig:triangulation}). 

In \cite{ABBL} an algorithm is described to associate a modular parameter to a triangulation of the torus. Basically, this is done by constructing a periodic embedding of the torus in the Euclidean plane which we require to be harmonic, by which we mean that any vertex is located at the centre of mass of its neighbours. Such an embedding is unique up to scaling, rotations, and translations if we insist that the sum of all squared edge lengths is minimized (figure 1). 

\begin{figure}
\begin{minipage}[b]{7.5cm}
\begin{center}
\includegraphics[width=6cm]{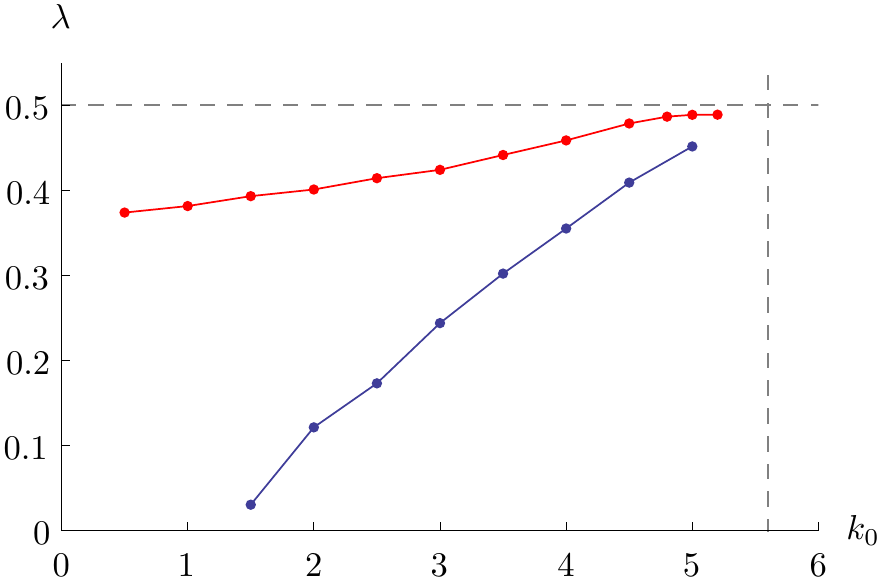}
\end{center}
\caption{Determined values of $\lambda$ from the minisuperspace method (in blue) and from extrinsic curvature (in red).\label{fig:lambdaplot}}\end{minipage}
\hspace{0.5cm}
\begin{minipage}[b]{7.5cm}
\begin{center}
\includegraphics[width=4cm]{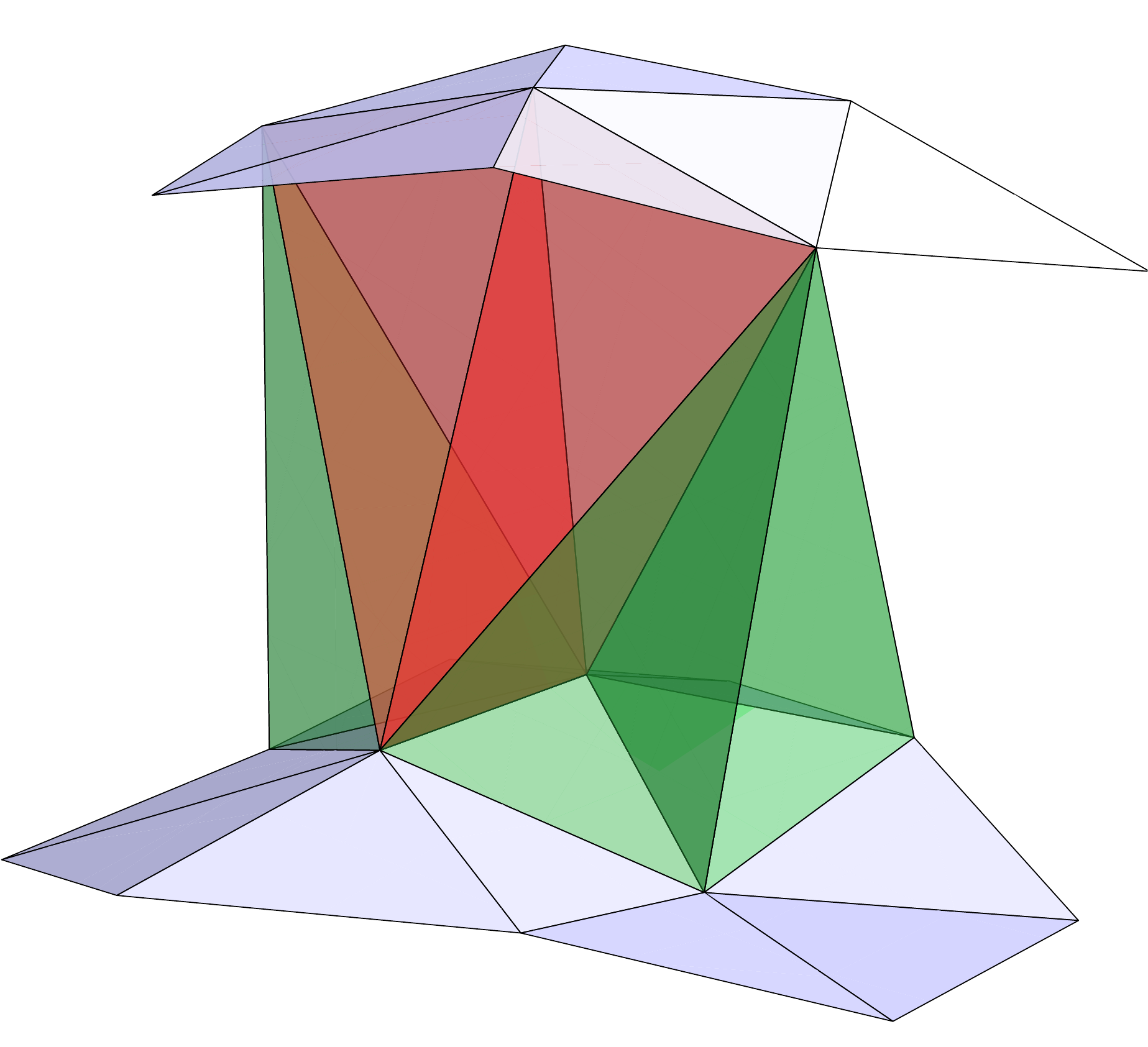}
\end{center}
\caption{The extrinsic curvature at an edge in the initial boundary is proportional to the number of tetrahedra connecting to it. \label{fig:extrinsic}}
\end{minipage}
\end{figure}

Let's see how this modulus appears in the effective action according to our ansatz. Restricting the kinetic term given by the modified Wheeler-DeWitt metric $\mathcal{G}_{\lambda}$ to the spatial volume $V$ and the modulus $\tau=\tau_1+i\tau_2$ we obtain
\begin{equation}
\kappa\int dt \left((\frac{1}{2}-\lambda)\frac{\dot{V}^2}{V} + \frac{1}{2A[g]}\frac{\dot{\tau}_1^2+\dot{\tau}_2^2}{\tau_2^2}\right), \quad\mathrm{where}\quad A[g] = \frac{\int d^2x\sqrt{g}\exp(2\Delta^{-1}R)}{\left(\int d^2x\sqrt{g}\exp(\Delta^{-1}R)\right)^2}.
\end{equation}
In a CDT simulation we can perform direct measurements of the functional $A[g]$ on the spatial triangulation at time $t$, the result of which is shown in black in figure \ref{fig:cinv} for several different boundary conditions (which we will not discuss here). However, we can also deduce the ratio $(1/2-\lambda)A[g]$ of the prefactors in the kinetic term by measuring the correlation functions $\langle V(t)V(t+\Delta t)\rangle$ and $\langle \tau_i(t)\tau_j(t+\Delta t)\rangle$ for small $\Delta t$. The results are shown in orange and red in figure \ref{fig:cinv}, where we have determined (for all four graphs simultaneously) the overall factor $1/2-\lambda$ by a best fit. We see that the overlap is reasonably good, which is in favour of our ansatz. This way we have determined $\lambda$ for various values of the CDT coupling $k_0$ as is shown in blue in figure \ref{fig:lambdaplot}.

\section{Extrinsic curvature at a fixed boundary}

Another more local way of probing the kinetic term is by studying the geometry near a fixed boundary. It can be shown that the fluctuations of the spatial metric $g_{ab}(\Delta t)$ a small time $\Delta t$ after a fixed initial two-dimensional surface are governed by the inverse $\mathcal{G}_{abcd}^{\lambda}$ of the modified Wheeler-De Witt metric in (\ref{eq:modifiedWdW}),
\begin{equation}\label{eq:ansatzmetricfluc}
\langle g_{ab}(\Delta t,x) g_{cd}(\Delta t,y) \rangle - \langle g_{ab}(\Delta t,x)\rangle \langle g_{cd}(\Delta t,y)\rangle \propto \delta(x-y) \mathcal{G}_{abcd}^{\lambda}.
\end{equation}
The natural way to test this ansatz would be to consider three-dimensional CDT configurations of which we fix the first spatial triangulation and observe how the geometry of the second spatial triangulation fluctuates. However, defining observables that capture these fluctuations and relating them to the continuum metric turns out to be difficult and highly ambiguous. Therefore, we make the assumption that the geometry of the second spatial triangulation is to large extent captured indirectly by the extrinsic curvature at the initial surface. Accordingly we test the ansatz (\ref{eq:ansatzmetricfluc}) in which we replace $g_{ab}(\Delta t)$ by the extrinsic curvature $K_{ab}$, but have to keep in mind that this substitution might have a systematic effect on $\lambda$. 

Viewing the CDT configuration as a piece-wise linear manifold, the extrinsic curvature has support on the edges and is (up to a constant) proportional to the number $N(e)$ of tetrahedra connecting to the edge $e$ (figure \ref{fig:extrinsic}). We have checked numerically that the fluctuations in $N(e)$ only correlate locally and the correlation does not depend on the system size. This means that if we send the system size to infinity only an ultralocal correlation will survive. We take our boundary to be a regular triangulation of the torus, because this allows us to unambiguously extract the ultralocal part of the correlations in $N(e)$. We have determined $\lambda$ by comparing this to our ansatz for the correlation of the extrinsic curvature. The results are shown in red in figure \ref{fig:lambdaplot}.

\section{Discussion}
At an effective level comparing CDT to the Euclidean Einstein-Hilbert action is problematic due to the conformal mode problem. However, if we replace its kinetic term by a non-covariant but positive-definite one given by  $\mathcal{G}_{\lambda}$ for $\lambda<1/2$ we obtain a sensible ansatz and we find qualitative agreement with the experiments described. These findings seem to strengthen an earlier observed connection between CDT and Ho\v{r}ava-Lifshitz gravity in \cite{AGJJL}.

The discrepancy in figure \ref{fig:lambdaplot} between the results of both methods indicates that the determination of $\lambda$ is quite subtle and probably sensitive to the precise ensemble of triangulations used. However, the overall behaviour is clear: as we approach the phase transition in CDT, corresponding to a critical coupling $k_0\approx 5.6$, $\lambda$ increases to $1/2$ at which point $\mathcal{G}_{\lambda}$ becomes degenerate.

A more detailed exposition of the presented measurements will appear soon.


\section{References}
\begin{thebibliography}{9}
\bibitem{AJL} Ambj\o{}rn J, Jurkiewicz J and Loll R 2001 {\it Phys. Rev.} D {\bf 64} 044011 ({\it Preprint} arXiv:hep-th/0011276)
\bibitem{AGJL} Ambj\o{}rn J, G\"orlich A, Jurkiewicz J and Loll R 2008 {\it Phys. Rev. Lett.} {\bf 100} 091304 ({\it Preprint} arXiv:0712.2485)
\bibitem{H} Ho\v{r}ava P 2009 {\it Phys. Rev.} D {\bf 79} 084008 ({\it Preprint} arXiv:0901.3775)
\bibitem{ABBL} Ambj\o{}rn J, Barkley J and Budd T G 2011 {\it Preprint} arXiv:1110.4649
\bibitem{AGJJL} Ambj\o{}rn J, G\"orlich A, Jordan S, Jurkiewicz J and Loll R 2010 {\it Phys. Lett.} B {\bf 690} 413-419 ({\it Preprint} arXiv:1002.3298)
\end{thebibliography}
\end{document}